# Spectral causality and the scattering of waves

ZEKI HAYRAN, AOBO CHEN, AND FRANCESCO MONTICONE *

*School of Electrical and Computer Engineering, Cornell University, Ithaca, New York 14853, USA*
* *francesco.monticone@cornell.edu*

**Abstract:** Causality – the principle stating that the output of a system cannot temporally precede the input – is a universal property of nature. Here, we show that analogous input-output relations can also be realized in the spectral domain by leveraging the peculiar properties of time-modulated non-Hermitian photonic systems. Specifically, we uncover the existence of a broad class of complex time-modulated metamaterials which obey the time-domain equivalent of the well-established frequency-domain Kramers-Kronig relations (a direct consequence of causality). We find that, in the scattering response of such time-modulated systems, the output frequencies are inherently prohibited from spectrally preceding the input frequencies, hence we refer to these systems as 'spectrally causal'. We explore the consequences of this newly introduced concept for several relevant applications, including broadband perfect absorption, temporal cloaking of an 'event', and truly unidirectional propagation along a synthetic dimension. By emulating the concept of causality in the spectral domain and providing new tools to extend the field of temporally modulated metamaterials ("chrono-metamaterials") into the complex realm, our findings may open unexplored opportunities and enable relevant technological advances in various areas of photonics and, more broadly, of wave physics and engineering.

## 1. Introduction

In 1958, F. R. Morgenthaler noted that light experiences frequency scattering upon encountering a temporal perturbation [1], analogous to the occurrence of wavevector scattering when light impinges on a spatial perturbation. Although the field of temporal/frequency scattering has received comparatively less attention than its spatial/wavevector counterpart [2],[3], several subsequent studies have unveiled the rich physics embodied by wave propagation and scattering through time-varying media [4],[5],[6]. More recent studies on dynamical optical systems have further demonstrated the unique capabilities of such systems in several key areas, such as for optical isolation [7],[8], delay-line buffering [9], topological phase transitions [10], among many others [11],[12],[13],[14],[15] thereby renewing interest in this field. With a few exceptions discussed in the following [16],[17],[18],[19],[20], however, such systems have been investigated under the assumption that the temporal perturbation is essentially Hermitian (*i.e.* it does not directly involve the loss or gain properties of the system). However, to exploit all the degrees of freedom and the full capabilities of an electromagnetic/photonic system, the extension of the relevant optical parameters to the complex domain is clearly essential, both in the context of conventional absorption/emission and gain processes, and for more exotic phenomena involving, for example, exceptional points [21]. Indeed, in the context of spatial scattering, parity-time (PT-) symmetric [22] or generalized non-Hermitian spatially modulated [23],[24] systems have been shown to exhibit many optical effects that are unattainable or difficult to achieve with their Hermitian counterparts, including unidirectional invisibility [25], exceptional-point based sensing and lasing [21], among others [22].

Temporally modulated systems might naturally lead to time-varying loss or gain, even when the designed modulation only involves the "Hermitian properties" of the system, e.g., the refractive index. For instance, this could be due to the opening of a radiation/absorption channel by the modulation, or direct energy pumping into or from the optical system as in the process of parametric amplification [16]. Instead, despite the unexplored potential of this research direction, only few studies [16],[17],[18],[19],[20] have focused on the effect of temporal non-Hermiticity, in which, for example, a refractive index modulation is deliberately accompanied by a specific modulation of the loss/gain coefficient. In these earlier studies, however, the modulations were strictly implemented under certain symmetry requirements, for example PT-symmetry as in Refs. [16],[17],[19]. Aside from leading to the stringent necessity of optical gain, such strict symmetry conditions may not be suitable in certain practical applications, where different (non-periodic and asymmetric) forms of modulation may be preferable and easier to implement. Moreover, other recent studies have focused on non-Hermitian systems with amplitude or amplitude-phase temporal modulations applied to guided waves in ring resonators [26],[27]. However, we note that these studies did not consider simultaneous temporal modulations of the real and imaginary part of the material constitutive parameters; instead, they incorporated additional modulating devices based on purely Hermitian modulations of the refractive index (e.g., Mach-Zehnder-interferometer based modulators). In this work, instead, we focus on the fundamental temporal/frequency scattering process in materials with direct modulations of both real and imaginary part of their susceptibility.

Specifically, here we uncover the existence of a broad class of non-Hermitian temporal modulations based on more general symmetries and relations between the real and imaginary parts of the material modulation function. We focus on temporal modulations that satisfy certain integral relations that are the temporal equivalent of the well-established Kramers-Kronig relations, and, thereby, possess a 'causal spectral response', *i.e.* the output frequencies generated by the temporal scattering process do not spectrally precede the input frequencies. To demonstrate the potential of this idea, we then present several representative applications, where the proposed concept of 'spectral causality' can have far reaching implications for the design of novel photonic devices. We note that unlike PT-symmetry requirements, the temporal Kramers-Kronig relations that lie at the basis of the concept of spectral causality do not involve such strict symmetry constraints as a perfectly balanced distribution of loss and gain; instead, they provide a way to design the imaginary part of the temporal modulation from the real part, or vice versa, regardless of their specific symmetries, with the goal to achieve certain advanced functionalities (perfect broadband absorption, invisibility, nonreciprocity, etc.). Hence, the ideas put forward in this article lead to a broader class of dynamically modulated structures with predetermined scattering properties, and they represent a new step toward the goal of rigorously extending the field of metamaterials into the temporal domain.

## 2. Theory

Consider a generic electromagnetic wave propagating in a homogeneous medium whose optical properties are dynamically varied in time, as illustrated in Fig. 1(a). We neglect magnetic effects and assume here that the medium exhibits only an electric polarization in response to an applied electric field. The constitutive relation between the electric field **E** and the electric displacement field **D** in a medium with a time-varying electric susceptibility can be written, in the general case, as $\mathbf{D}(\mathbf{r},t) = \varepsilon_0 \left[ \varepsilon_s \mathbf{E}(\mathbf{r},t) + \int dt' \chi(t,t') \mathbf{E}(\mathbf{r},t-t') \right]$, where $\chi(t,t')$ is the time-dispersive time-varying electric susceptibility, $\varepsilon_s$ is the static and spatially-homogenous background relative permittivity, and $\varepsilon_0$ is the permittivity of free space [28],[29]. The function $\chi(t,t')$ can be interpreted as a dispersive susceptibility $\chi(t')$ (material impulse response), which varies with time $t$. Here, as illustrated in Fig. 1(d), the time variable $t'$ (and its Fourier-pair frequency variable $\omega'$) primarily encode the dispersion properties of the material (the fact that the material response is not instantaneous), whereas the time variable $t$

(and its Fourier-pair frequency variable $\omega$) primarily encode the time-modulation properties imparted by an external modulation mechanism (as commonly done in the literature [28],[29]]). In this regard, $\chi(t,t')$ corresponds to a time-varying susceptibility function capturing both the dispersion properties of the medium and how the external modulation changes these properties. While this formalism is general, to simplify the analysis, it is usually assumed that the temporal modulation occurs on a time scale larger than the response time of the material. In other words, under this approximation, we operate at steady-state with a time-harmonic wave of frequency $\omega'$, and the material is then modulated in $t$-time on a time scale larger than the time it takes the material to approximately reach a new steady state (for non-adiabatically modulated dispersive susceptibility models see also Refs. [29],[30],[31]). In addition, it is important to note that, if we consider the material response in $\omega'$-frequency domain, the function $\chi(t,\omega')$ is complex, with the imaginary part corresponding to time-varying loss or gain. In the following, we first consider a non-dispersive material, such that $\chi(t,\omega') = \chi(t)$ (i.e., an instantaneous time-varying material response $\chi(t,t') = \chi(t)\delta(t')$) and will then discuss the role of dispersion at the end of this section.

The vector wave equation in time domain, in a non-magnetic material with no sources, reads

$$\nabla \times \nabla \times \frac{\mathbf{E}(\mathbf{r},t)}{\mu_0} + \frac{\partial^2 \mathbf{D}(\mathbf{r},t)}{\partial t^2} = 0, \tag{1}$$

where $\mu_0$ is the permeability of free space. Since a time-varying spatially homogeneous medium conserves the wavevector $\mathbf{k}$ of a propagating wave, we write the electric field in the medium in plane-wave form: $\mathbf{E}(\mathbf{r},t) = \mathrm{Re}\left[\mathrm{E}(t)\exp(-i\mathbf{k}\mathbf{r})\right]\mathbf{n}$, where $\mathbf{n}$ is the polarization vector, while the function $\mathrm{E}(t)$ needs to be determined ($\mathrm{E}(t)$ is a complex phasor-like quantity, which, in the time-invariant time-harmonic case, is simply given by $\mathrm{E}(t) = E_0 e^{+i\omega t}$, where $E_0$ is a complex constant). The displacement field can similarly be written as $\mathbf{D}(\mathbf{r},t) = \mathrm{Re}\left[\mathrm{D}(t)\exp(-i\mathbf{k}\mathbf{r})\right]\mathbf{n}$. Using the wave equation, the constitutive relation in the dispersionless case, and $|\mathbf{k}| = \omega_i\sqrt{\mu_0\varepsilon_0\varepsilon_s}$, we find the differential equation that governs the time evolution of $\mathrm{D}(t)$ as

$$\left(\varepsilon_s \frac{d^2}{dt^2} + \varepsilon_s \omega_i^2\right)\mathrm{D}(t) = -\chi(t)\frac{d^2}{dt^2}\mathrm{D}(t), \tag{2}$$

where $\omega_i$ is the "incident" frequency (namely, the initial steady-state $\omega'$-frequency of the field in the material before the modulation starts). Here, we define the Green's function for the operator on the left side of Eq. (2) as $G(t)$, leading to the following recursive expression for $\mathrm{D}(t)$,

$$\mathrm{D}(t) = D_0 e^{i\omega_i t} + \int d\tau\, G(t-\tau)\bigl(-\chi(\tau)\bigr)\frac{d^2\mathrm{D}(\tau)}{d\tau^2}, \tag{3}$$

where $D_0$ is the incident amplitude of $\mathbf{D}$. Fourier transforming Eq. (3) (going from $t$-time domain to $\omega$-frequency domain) and reformulating it in terms of a scattering series gives (see Supplement 1), we obtain

$$\tilde{D}^{(N)}(\omega) = 2\pi D_0 \delta(\omega - \omega_i) + \sum_{m=1}^{N} \tilde{D}_s^{(m)}(\omega), \tag{4}$$

with,

$$\tilde{D}_s^{(m)}(\omega) = \tilde{G}(\omega)\left[\left(\tilde{\chi} * \omega^2 \tilde{D}_s^{(m-1)}\right)(\omega)\right] \quad \text{for } m > 1, \tag{5}$$

$$\tilde{D}_s^{(1)}(\omega) = 2\pi \omega_i^2 D_0 \tilde{G}(\omega)\left[(\tilde{\chi} * \delta)(\omega - \omega_i)\right], \tag{6}$$

where the operator "$*$" denotes convolution. In the above series, $\tilde{D}_s^{(1)}(\omega)$ can be regarded as the temporal analogue of the first-order Born approximation in the theory of spatial scattering [32]. Notice that in the case of a time-independent $\chi(t) = \chi$ (*i.e.* in the absence of any temporal modulation), $\tilde{\chi}(\omega)$ becomes $\tilde{\chi}(\omega) = 2\pi \chi \delta(\omega)$, and the convolution simply results in $(\tilde{\chi} * \delta)(\omega - \omega_i) = 2\pi \chi \delta(\omega - \omega_i)$ in Eq. (6). Hence, no spectral change to the incident field is expected as a first order approximation. Moreover, upon investigating all higher-order terms of the scattering series, $\tilde{D}_s^{(m)}$ in Eq. (5), we see that the incident field does not experience any spectral change at all (since throughout the series the only convolution operation is with $\tilde{\chi}(\omega)$), which is expected due to linearity and time-invariance. On the other hand, if $\chi$ becomes time-dependent, the convolution in Eq. (6) produces a term $\tilde{D}_s^{(1)}(\omega)$ that is no longer a delta function centered at $\omega_i$. In other words, the time-varying susceptibility "couple" the incident frequency $\omega_i$ with different frequency components (*i.e.*, it generates different frequencies) depending on the spectral content of $\tilde{\chi}(\omega)$. Notably, an arbitrary $\tilde{\chi}(\omega)$ typically couples the incident frequency with 'negative' frequencies, analogous to the creation of negative wavevectors that propagate backwards in space in the context of spatial scattering. However, the notion of negative frequencies does not imply backward propagation in time, as this would obviously violate temporal causality. Instead, the coupling to negative frequencies should be interpreted as the generation of phase-conjugated waves that propagate backwards in space, leading to the phenomenon of temporal reflection [33] (see Fig. 1(b)).

Interestingly, in the special case when $\tilde{\chi}(\omega < 0) = 0$, the modulation function $\tilde{\chi}(\omega)$ cannot promote frequency down-conversion, thereby the generation of any lower (including negative) frequencies is prohibited at all orders (see Fig. 1(c)). This can be seen as the spectral analogue of the concept of temporal causality, which states that a causal response function $\text{H}(t)$, for which $\text{H}(t < 0) = 0$, applied to a generic input function $\text{f}(t)$, for which $\text{f}(t < t_0) = 0$, results in an output function $\text{g}(t) = (\text{H} * \text{f})(t)$ for which $\text{g}(t < t_0) = 0$. In other words, the output of a causal system cannot temporally precede the input. By analogy, we thus refer to any spectral susceptibility profile that satisfies $\tilde{\chi}(\omega < 0) = 0$ as '*spectrally causal*', implying that the scattered waves arising due to such a temporal modulation profile cannot have components at frequencies that spectrally precede the input frequencies.

Since, by definition, a spectrally causal $\tilde{\chi}(\omega)$ is asymmetric around the frequency-axis origin, the temporal profile $\chi(t)$ is necessarily complex due to the properties of the Fourier transform (recall that a complex temporal profile is allowed here since we are considering the material response in $\omega'$-frequency domain). Therefore, one can expect that loss and/or gain must accompany the temporal modulation profile $\chi(t)$ to ensure its spectral causality, similar to the implications of the well-established Kramers-Kronig relations for a temporally casual response function (the real and imaginary part of the Fourier transform of a causal square-integrable function are not independent, and their mutual dependence is expressed by Kramers-Kronig relations). We can formalize this analogy by using the analytical properties of the Fourier transform of the causal function $\tilde{\chi}(\omega)$, that is, $\chi(t)$, to derive relations, analogous to the standard Kramers-Kronig relations, between the real and imaginary parts of the temporal modulation profile.

In fact, by noting that a spectrally casual $\chi(t)$ is an analytic function in the upper complex time plane if $\tilde{\chi}(\omega)$ and $\chi(t)$ are square integrable on the real line (which is satisfied by considering a finite temporal modulation that vanishes at large times), we can derive integral relations for $\chi(t)$ that relate its real, $\chi_{re}(t)$, and imaginary, $\chi_{im}(t)$, parts using complex analysis [34]:

$$\chi_{re}(t) = -\frac{1}{\pi} P \int_{-\infty}^{\infty} dT \frac{\chi_{im}(T)}{t-T}, \tag{7}$$

$$\chi_{im}(t) = \frac{1}{\pi} P \int_{-\infty}^{\infty} dT \frac{\chi_{re}(T)}{t-T}, \tag{8}$$

where P indicates the Cauchy principal value of the integral. The form of Eqs. (7) and (8) differs from the conventional Kramers-Kronig relations by a minus sign (originating from the fact that the conventional Kramers-Kronig relations are based on analyticity in the *lower* complex frequency plane for the chosen conventions). Hence, we refer to them as the *anti-Kramers-Kronig relations* in time domain. Such relations are a key result of this work, as they allow one to choose the specific temporal modulation function of the real and imaginary part of the susceptibility (hence, the temporal modulation of the refractive index and of the loss/gain coefficient) such that the modulation is spectrally causal and, therefore, it does not induce frequencies preceding the incident frequency. As discussed in the following, this is potentially relevant for a plethora of applications. Importantly, while we assumed causality in the spectral domain to derive Eqs. (7) and (8), one can invoke Titchmarsh Theorem [35],[36] to demonstrate that the converse is also true, namely, that Eqs. (7) and (8) imply causality in the spectral domain.

A rather important point to stress here is that additional constraints must be applied on a physical susceptibility function $\chi(t,\omega')$ such that *temporal* causality is not violated. Particularly, the conventional Kramers-Kronig relations imply that (i) the complex $\omega'$-frequency-domain susceptibility $\chi(t,\omega')$ of any physically realizable material except vacuum is always dispersive, namely, it is not a constant with respect to the frequency $\omega'$ of a time-harmonic wave at any *t*-time instant, and (ii) frequency dispersion is intimately related to absorption/gain processes.

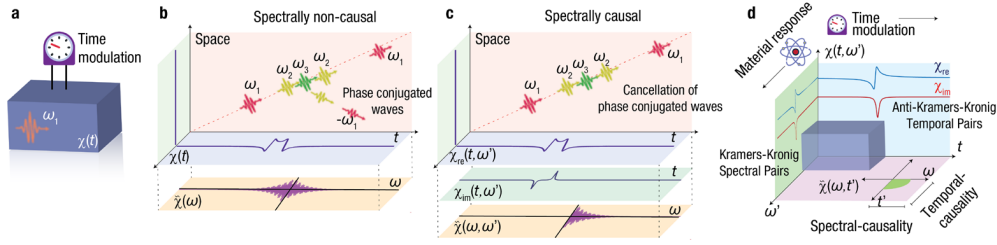

**Fig. 1.** Spectral causality and its physical implications. (a) Propagation of a quasi monochromatic electromagnetic wave within a homogeneous medium whose optical parameters can be controlled dynamically in time. (b) Upon encountering an arbitrary, real, temporal electric susceptibility perturbation $\chi(t)$, the pulse will experience time-refraction and time-reflection leading, respectively, to time-shift (delay) in the transmitted field and the generation of negative frequencies (phase-conjugated and reflected waves). Bottom panel: Fourier transform of $\chi(t)$. (c) A complex susceptibility perturbation, on the other hand, can prevent the generation of any negative frequencies if the real and the imaginary parts of the time-varying electric susceptibility function are anti-Kramers-Kronig temporal pairs. Bottom panel: Fourier transform of $\chi(t,\omega')$; the modulation is spectrally causal since $\tilde{\chi}(\omega<0,\omega')=0$. (d) For physical realizability and spectral causality, $\tilde{\chi}(\omega,\omega')$ must be dispersive and satisfy the anti- and conventional Kramers-Kronig relations in the time and frequency domain, respectively, as discussed in the text. Red dotted lines in (b) and (c) denote the propagation trajectory of the incoming wave in the medium in the absence of any time modulation.

Hence, the above discussion for the case of spectral causality holds true and is physically realizable only if the following conditions are met (illustrated in Fig. 1(d)): (1) for every value of $\omega'$ that is being generated throughout the frequency scattering process, the condition $\tilde{\chi}(\omega<0,\omega')=0$ must be satisfied to ensure spectral causality; (2) at every $t$-time instant, $\chi(t,\omega')$ must satisfy the conventional Kramers-Kronig relations along the $\omega'$-axis to ensure temporal causality (see Fig. 1(d)). As mentioned above, frequency dispersion is a consequence of temporal causality and must be included in any model of a polychromatic system to ensure its physical nature (all the full-wave simulations of time-varying systems in the next sections properly include frequency dispersion).

In light of these considerations, we conclude that a temporally modulated material susceptibility that satisfies both the anti-Kramers-Kronig relations (as in Eqs. (7) and (8)) in the $t$-time domain, at least for a certain $\omega'$-bandwidth, and the conventional Kramers-Kronig relations in the $\omega'$-frequency domain for all $t$-time instants, is spectrally casual and physically realizable (Fig. 1(d)). These conditions will ensure that an incoming wave only scatters to higher frequencies with no "back-scattering" to lower ones (including negative frequencies, corresponding to phase-conjugated and reflected waves). In addition, since the temporal part of **D** is separated from its spatial part, the spectral causality property will be independent of the spatial properties of the incoming wave, such as polarization, beam-shape, and propagation direction [37]. Also important to point out is that the anti-Kramers-Kronig relations are sufficient but not necessary to realize a reflectionless time-varying system. Indeed, purely Hermitian temporal profiles can also be found that are reflection-free [37],[38], which, however, do not provide the same flexibility in terms of controlling the frequency scattering process, as further elucidated in the next sections.

Finally, while the considered modulations are spatially homogeneous, namely, they do not vary in space (unlike, for example, traveling-wave space-time modulations [7],[8]), and wave propagation in homogenous media was assumed in the simplified derivations above, the concept of spectral causality and the temporal Kramers-Kronig relations do not require the entire system to be spatially homogeneous. Indeed, as shown in Sections 3.1 and 3.3, the implications of spectral causality hold also in configurations with spatial interfaces owing to the finiteness of the temporal modulation (*i.e.*, the incident propagating pulse needs to experience the modulation only while the susceptibility perturbation is nonnegligible, hence the system can be truncated within spatial boundaries that are "synchronized" with the temporal boundaries of the modulation). Most examples of applications discussed in the following indeed involve structures that are spatially inhomogeneous with only some spatial regions that are temporally modulated. Moreover, Section 3.3 extends the theory developed here to explicitly include spatially inhomogeneous media.

## 3. Applications

### 3.1 Broadband reflectionless spectrally causal absorbers

One immediate application to verify our theory is the design of a spectrally casual absorber, namely, a device that automatically prevents the coupling of the incident frequency to any negative frequencies and, therefore, eliminates any back-reflections and fully absorbs the incident wave, as illustrated in Fig. 2(a). To satisfy both the temporal anti-Kramers-Kronig relations and the conventional spectral Kramers-Kronig relations for such an absorbing medium, we propose the following material model,

$$\chi(t,\omega') = \frac{\omega_p^2}{\left(\omega_0 - K(t - t_{\text{offset}})\right)^2 - \omega'^2 + i\gamma\omega'}, \tag{9}$$

which is a standard Lorentz-type dispersion with plasma frequency $\omega_p$, damping frequency $\gamma$, and with a time-varying resonance frequency $\omega'_0(t) = |\omega_0 - K(t - t_{\text{offset}})|$ (see lower right inset of Fig. 2(a)). Here, the parameter $K$ determines the rate of change of the resonance frequency and $t_{\text{offset}}$ provides a $t$-time offset. Eq. (9) is plotted on the $(t,\omega')$ plane in Figs. 2(b) and 2(c), for parameters given in Supplement 1. It is apparent that Eq. (9) fully satisfies the conventional Kramers-Kronig relations along the $\omega'$-frequency axis at any time instant $t$. Moreover, it also approximately satisfies the anti-Kramers-Kronig relations along the $t$-axis within a certain time window that can be chosen via $t_{\text{offset}}$ (see Supplement 1). Note that this spectrally causal non-Hermitian temporal modulation only involves loss, while gain is not required.

Fig. 2(d) shows the spectral change of an incident broadband wave obtained through fully causal time-domain simulations (see Supplement 1 for additional details). As expected, one can observe the absence of any reflections *at any frequency* (absence of standing wave behavior) and the frequency upshifting of the wave within the material. We stress here that since the material is physical and temporally causal, the wave experiences a different $\chi(t,\omega')$ profile as the frequency $\omega'$ upshifts. In other words, the temporal profile that the incident wave ultimately experiences is a synthesis of various $\chi(t,\omega')$ profiles determined by the frequency scattering process. However, since the material dispersion, Eq. (9), is designed to satisfy spectral causality also at these scattered higher frequencies (as shown in Figs. 2(b) and 2(c)), no lower frequencies (including negative frequencies) are generated throughout the propagation and scattering process. Hence, the wave is fully absorbed within the lossy medium without any back scattering, as seen in Fig. 2(d). Remarkably, the implications of our simplified theory above still hold, and no reflection occurs, even though the temporal variation of $\chi(t,\omega')$ is relatively fast and non-adiabatic (the temporal linewidth of the Lorentzian loss profile in Fig. 2(c) is comparable to the period of the incoming wave). In other words, although our assumption that the temporal modulation is applied on a time scale much larger than the response time of the material may not be valid in this scenario, the qualitative predictions of our theory still hold even for these relatively fast temporal variations. This is consistent with the recent findings in Ref. [39] that, even if the temporal variation is on a scale comparable to the period of the driving wave, one can still define a physically meaningful time-varying susceptibility $\chi(t,\omega')$. However, we note that more research is needed to understand the full ramifications of non-adiabaticity in the context of spectrally causal temporal modulations.

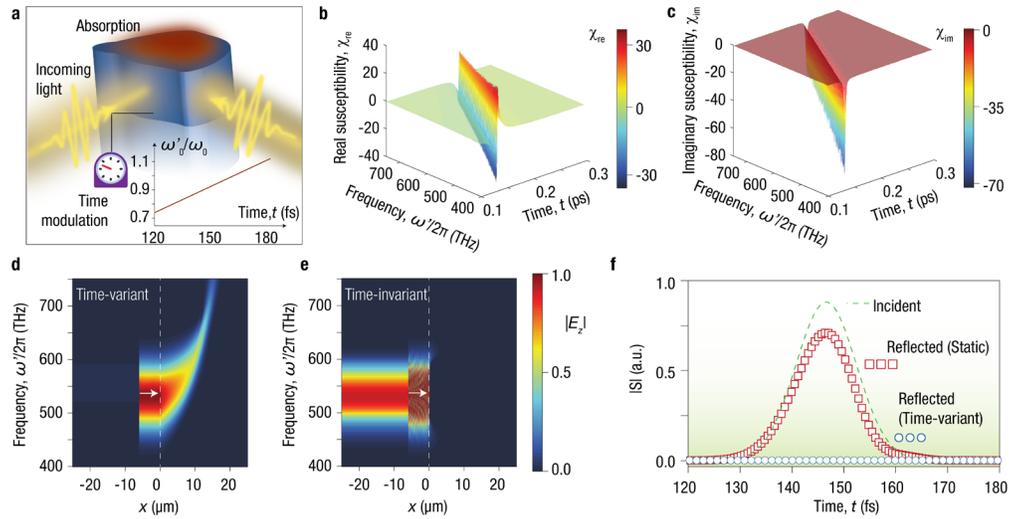

**Fig. 2.** Broadband reflectionless absorption induced by spectral causality. (a) The absence of negative frequencies in a spectrally-causal time-scattering process can be exploited to achieve broadband omnidirectional perfect absorption in a lossy medium. The lower right inset shows the modulation of resonant frequency $\omega_0$ in $t$-time. Real (b) and imaginary (c) parts of the $\omega'$-*frequency-dispersive* and $t$-*time-varying* electric susceptibility of the designed spectrally (and temporally) causal lossy Lorentz medium. (d,e) Spectrally resolved electric field magnitude along the propagation direction for the time modulated (d) and static (time-invariant) (e) absorbing medium. (f) Amplitude of the reflected Poynting vector recorded in time at the spatial position $x = -25$ μm. The green dashed line shows the envelope of the instantaneous Poynting vector of the input pulse. In (d) and (e) white arrows denote the excitation direction, and the white dashed lines mark the interface between free space and the absorbing medium. The details of these full-wave simulations and the modulation parameters are given in Supplement 1.

For comparison, a static structure with the same Lorentz dispersion as in Eq. (9) but without the temporal variation in $\omega_0$ (i.e., $K = 0$) was also analyzed and the results are provided in Fig. 2(e). We see that despite the high losses in the medium, the wave is mostly backscattered rather than absorbed due to the high impedance mismatch resulting from the large amplitude of the Lorentzian profile. This is a rather striking result showing how a suitable spectrally causal temporal modulation can lead to nearly perfect *broadband absorption* even when the corresponding static structure is largely impedance mismatched and highly reflective. Fig. 2(f) provides a further quantitative comparison between the reflected Poynting vector amplitude |S| (flow of energy per unit area per unit time) in the time-varying spectrally causal case and the time-invariant case. The difference between the reflected |S| and the input |S| quantifies the absorbed energy within the system, which confirms that, in contrast with the static structure, the designed dynamic medium absorbs nearly the entire broadband input radiation (see also Visualization 1 (†), which provides a detailed view into the broadband pulse propagation for both the dynamic and the static case).

Another rather important point worth mentioning is that reflectionless absorption over a broad bandwidth does not necessarily require the temporal modulation of the system parameters. As discussed in, for example, Refs. [40],[41],[42], broadband reflectionless absorption can be achieved in certain time-invariant and linear structures, such as reciprocal adiabatically tapered waveguides [40], or nonreciprocal (albeit time-invariant) terminated one-way waveguides [41], [42]. Here, instead of relying on spatial tapering or inherently unidirectional propagation, the lack of reflections is enabled by the spectral causality of the material (absence of scattered frequencies preceding the input frequency), thus enabling a potentially more compact and flexible system to achieve perfect broadband absorption, without the need for tapering or strictly unidirectional modes, and independently of the polarization, beam shape, and propagation direction of the incoming wave.

### 3.2 Temporal cloaking of an 'event'

As mentioned above, an arbitrary time-varying $\chi(t,\omega')$ profile can induce frequency scattering, including back-reflected waves. An external observer will then be able to tell the existence of the time-varying perturbation by simply detecting any scattered fields (see Fig. 3(a)). To illustrate this scenario, Figs. 3(b) and 3(c) show, respectively, a refractive index perturbation and the space-time evolution of a pulse propagating through a medium with such a perturbation. The reflections induced by the perturbation can be clearly seen in Fig. 3(c), in the form of fields propagating forward in $t$-time but in the backward spatial direction. In this regard, another intriguing application of the proposed concept of spectral causality is to design a temporal 'cloak' that reduces any back- and forward- scattered waves induced by a refractive index perturbation, restoring the incident wave after a certain time (analogous to spatial cloaks which restore the incident field after a certain *distance*), as shown in Fig. 3(d).

To get more insight into this process, we first consider a temporal profile $\chi(t,\omega')$ written as the sum of two generic profiles of the same type but with different weight and temporal offset, $\chi(t,\omega') = f_1 \chi_0(t - t_{\text{offset},1}, \omega') + f_2 \chi_0(t - t_{\text{offset},2}, \omega')$. Fourier transforming $\chi(t,\omega')$ gives

$$\tilde{\chi}(\omega,\omega') = 2if_2 \tilde{\chi}_0(\omega,\omega') \sin\left(\omega \frac{t_{\text{offset},1} - t_{\text{offset},2}}{2}\right) \exp(-i\omega \frac{t_{\text{offset},1} + t_{\text{offset},2}}{2}) + (f_1 + f_2)\tilde{\chi}_0(\omega,\omega')\exp(-i\omega t_{\text{offset},1}).$$
(10)

We assume that $\tilde{\chi}_0(\omega,\omega')$ (and therefore $\tilde{\chi}(\omega,\omega')$) is approximately spectrally causal, namely, it satisfies the anti-Kramers-Kronig relations within a certain $t$-time window (as the Lorentz medium presented in the previous section). Here, we are especially interested in suppressing the forward and backward scattered waves at the same frequency $\omega_i$ of the incident wave. Other frequency components of the scattered field would not be able to propagate in the same medium after the modulation ends, as there would not be any allowed propagation modes at that frequency $\neq \pm\omega_i$ and value of momentum (momentum of the propagating wave does not change since the system is spatially homogenous); these different frequency components would then be dissipated by any loss in the static system.

By referring back to Eq. (6) one notes that, as a first-order approximation, the forward- and backward-scattered waves at frequency $+\omega_i$ and $-\omega_i$, that is, $\tilde{D}_s^{(1)}(\pm\omega_i)$, are determined only by the values of the susceptibility function, $\tilde{\chi}(0,\omega')$ and $\tilde{\chi}(-2\omega_i,\omega')$, respectively (due to the convolution with $\delta(\omega-\omega_i)$). Thus, if these values (frequency components) of the modulation function vanish simultaneously in Eq. (10), the scattered waves $\tilde{D}_s^{(1)}(\pm\omega_i)$ will be suppressed completely. In this case, since any higher-order scattered term is directly related to the lower-order terms (see Eq. (5)), the *complete* suppression of the first-order contribution will, in fact, also translate into zero scattering at all orders, i.e., $\tilde{D}_s^{(m)}(\pm\omega_i)=0$. Thus, this would lead to an invisibility effect at frequency $\omega_i$. In Eq. (10), the condition $\tilde{\chi}(-2\omega_i,\omega')=0$ is automatically satisfied because of the spectral causality of $\tilde{\chi}_0(\omega,\omega')$. However, we note that $\tilde{\chi}_0(0,\omega')\neq 0$ because the considered temporal modulation profile satisfies the anti-Kramers-Kronig relations only within a certain $t$-time interval (due to this truncation, the integral $\tilde{\chi}_0(0,\omega')=\int\chi_0(t,\omega')dt$ will not vanish anymore). Thus, although the first term in Eq. (10) automatically vanishes for $\omega=0$, the second term vanishes only if $f_1=-f_2$, which can be satisfied considering a time-varying medium with a balanced gain-loss profile. Following this insight, we employ the following dispersion model (see Supplement 1) to realize an invisible ('cloaked') refractive-index perturbation,

$$\chi(t,\omega') = \frac{f_1\omega_p^2}{\left(\omega_0 - K(t-t_{\text{offset},1})\right)^2 - \omega'^2 + i\gamma\omega'} + \frac{-f_1\omega_p^2}{\left(\omega_0 - K(t-t_{\text{offset},2})\right)^2 - \omega'^2 + i\gamma\omega'}. \tag{11}$$

The "temporal cloak" profile consisting of the temporal loss and gain modulation is shown in Fig. 3(e), while Fig. 3(f) shows the space-time evolution of the pulse propagating within the temporally cloaked medium (see also Visualization 2 (†) for a time animation compared with the uncloaked case). Note the absence of any back-scattering from the perturbation (no fields propagating in the backward spatial direction), and the re-emergence of the absorbed wave after a certain $t$-time interval as a result of amplification. Additional scattered high-frequency components are still present in the system, which are, however, unable to propagate and are eventually absorbed, consistent with the discussion above (such spatially localized, temporally decaying fields can clearly be seen in Fig. 3(f)). Moreover, we note that the fields between the lossy and active space-time regions in Fig. 3(f) are not identically zero, such that the wave can be reconstructed through the amplification of these small fields. In other words, the system does not need prior knowledge about the incoming wave to reconstruct it by amplification after it passes through the lossy region, but it automatically provides a 'channel' between these two space-time regions that allows the reconstruction of the wave through the concerted action of loss and gain as guaranteed by spectral causality (this behavior can be considered the direct temporal equivalent of the PT-symmetric spatial cloaks studied in [25],[43], where an object is rendered invisible through the introduction of a suitable spatial distribution of loss and gain on its front and back, respectively). We acknowledge, however, that in practice this amplification and reconstruction process (in either the temporal or spatial cloaking case) would certainly add some noise to any signal being transmitted. Moreover, while the gain-enabled regeneration process requires no information about the propagating pulse, the overall cloaking process (as the absorption process in the previous section) still requires knowledge of the pulse arrival time.

This piece of information could then be provided through a passive mechanism, as suggested in [44] for different time-modulated systems, in which a switching element or phase transition is triggered by the arrival of the pulse, activating the spectrally causal modulation while the wave is propagating through the medium.

To emphasize the importance of the role played by spectral causality in this invisibility/cloaking effect, we have provided in Visualization 2 (†) an alternative case with balanced gain and loss where, however, the profile does not respect spectral causality. Although the gain and loss profile is identical as in the cloaked case, here the interplay between the real and the imaginary parts of the temporal modulation is broken (hence, Eq. (10) does not vanish for forward- and back-scattered waves). Thus, as seen in Visualization 2 (†), no cloaking effect occurs for such a non-causal profile, despite having a balanced gain-loss distribution.

We speculate that the proposed cloaking scheme may be useful for shielding information transfer from temporal perturbative effects (e.g. thermal or electrical fluctuations, mechanical movement/vibrations, etc.) that can affect the material properties of the medium, and the functionality of the system, as a function of time. We also stress here that, since the cloaking process requires gain materials, the stability of the system must be carefully analyzed to avoid any diverging oscillations [45],[46],[47]. A detailed stability analysis is beyond the scope of this paper. Nevertheless, we speculate that the proposed cloaking mechanism should be relatively robust against such instabilities. This is because no feedback mechanism exists that can lead to a resonance effect (due to the absence of any back-scattered waves) and, unlike static active systems, the gain materials here are *temporally* localized, *i.e.*, the gain process is not permanent, which may allow any unstable response to die out after gain is switched off. A detailed analysis of these effects will be the subject of future work.

Finally, we note that a different (albeit related) strategy to realize a cloaking effect could be envisioned by using a spectrally causal modulation to create a sort of 'gap' in the space-time response of the system (similar to the space-time response in Fig. 3(f) at around 0.25 ps), which could then be used to conceal certain objects and perturbations. Indeed, within this space-time window, invisibility could be achieved by overlapping the upshifted frequencies due to the spectrally causal modulation with a transparency window in the dispersion function of the object/perturbation, hence allowing the incident wave to tunnel through it with no reflections. Such transparency windows are commonly found, for instance, in plasmas or plasmonic materials, which can become nearly transparent above their plasma frequency. We are currently investigating such a cloaking strategy in more details as part of a future publication.

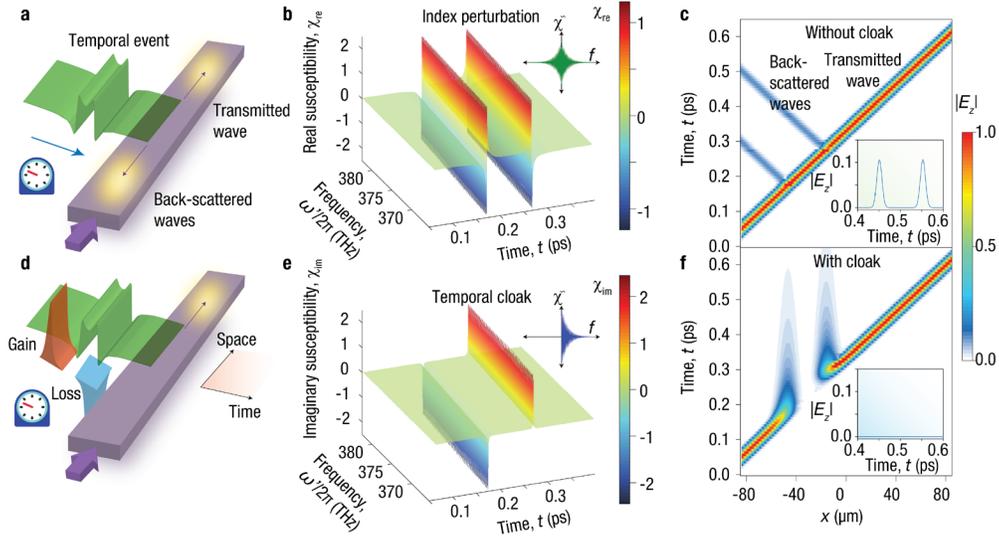

**Fig. 3.** Temporal cloaking of a refractive-index perturbation via an anti-Kramers-Kronig-pair cloak. (a) A temporal refractive-index modulation scatters an incident propagating wave, thereby revealing its existence to an observer. (b) The temporal perturbation consists of a purely Hermitian electric susceptibility modulation (a modulation of the refractive index). (c) Space-time evolution of a propagating pulse along the propagation direction in the presence of the temporal perturbation. (d,e) A temporal 'cloak' consisting of a gain-loss temporal modulation can be designed via the anti-Kramers-Kronig transformation of the refractive-index perturbation, as discussed in the text. (f) The temporal cloak effectively renders the perturbation 'invisible', hence restores the propagating wave and suppresses reflections. The lower right insets in (c) and (f) show the reflected electric field amplitude at $x = -85$ μm recorded in time. The details of these full-wave simulations and modulation parameters are given in Supplement 1.

### *3.3 Loss-induced unidirectional transport along a synthetic dimension*

Another intriguing application of the proposed concept of spectral causality is related to wave 'propagation' along a synthetic dimension (see Fig. 4(a)). Such synthetic dimensions in photonic structures have recently become the subject of intense research, as they enable observing higher-dimensional physics in seemingly lower-dimensional systems [48],[49],[50]. Here, we theoretically demonstrate that spectral causality can lead to *truly one-way propagation* along a synthetic dimension defined by the sequence of resonant frequencies of a resonator, as illustrated in Fig. 4(a). Since the different modes of a resonator have different momentum/wavevector components (spatial spectrum) [51], we revisit our theory to include also momentum coupling between different momentum/frequency-separated modes. We re-write Eq. (1) with a time- and space-dependent susceptibility $\chi(\mathbf{r},t,\omega') = \chi(\mathbf{r},t)$ (again assumed dispersionless for simplicity) as,

$$\nabla \times \nabla \times \frac{\mathbf{n}\mathrm{E}(\mathbf{r},t)}{\mu_0} + \mathbf{n}\frac{\partial^2 \varepsilon_s \mathrm{E}(\mathbf{r},t)}{\partial t^2} = -\mathbf{n}\frac{\partial^2 \mathrm{P}(\mathbf{r},t)}{\partial t^2}, \qquad (12)$$

$$\mathrm{P}(\mathbf{r},t) = \chi(\mathbf{r},t)\mathrm{E}(\mathbf{r},t),$$

where P is the polarization density generated by $\chi(\mathbf{r},t)$. Decomposing the harmonic components of $\chi(\mathbf{r},t)$ and assuming its temporal and spatial parts are separable gives,

$$\chi(\mathbf{r},t) = \frac{1}{(2\pi)^2} \int dk\, \tilde{\chi}(\mathbf{k}) e^{-i\mathbf{k}\mathbf{r}} \int d\omega\, \tilde{\chi}(\omega) e^{i\omega t}. \tag{13}$$

Then, we decompose E in its frequency eigenmodes, each expanded into its spatial harmonic components, obtaining

$$\mathrm{E}(\mathbf{r},t) = \frac{1}{2\pi} \sum_m c_m e^{i\omega_m t} \mathrm{E}_m(\mathbf{r}) = \frac{1}{2\pi} \sum_m c_m e^{i\omega_m t} \int dk\, \tilde{\mathrm{E}}_m(\mathbf{k}) e^{-i\mathbf{k}\mathbf{r}}, \tag{14}$$

where $c_m$ and $\omega_m$ are the amplitude and frequency of the $m$th mode, respectively. $\mathrm{E}_m(\mathbf{r})$ is the eigenmode spatial field profile that can be obtained by solving the eigenequation $\nabla \times \nabla \times \mathbf{n}\mathrm{E}_m = \mu_0 \varepsilon_s \omega_m^2 \mathbf{n}\mathrm{E}_m$. The induced P for a specific mode $m$ then becomes,

$$\mathrm{P}(\mathbf{r},t) = \frac{c_m}{(2\pi)^3} \int dk\, \left(\tilde{\mathrm{E}}_m(\mathbf{k}) * \tilde{\chi}(\mathbf{k})\right) e^{-i\mathbf{k}\mathbf{r}} \int d\omega\, \tilde{\chi}(\omega) e^{i(\omega_m+\omega)t}. \tag{15}$$

Eq. (15) shows that a spatially homogeneous $\chi(\mathbf{r},t)$ (hence, $\tilde{\chi}(\mathbf{k}) \propto \delta(\mathbf{k})$) cannot couple the different wavevector components of momentum-separated different modes (in other words, $\tilde{\chi}(\mathbf{k})$ generates an induced polarization current with zero overlap integral with the field distribution of a different mode, resulting in zero coupling). One way to break the spatial symmetry is to apply the temporal modulation on limited portion of the resonator, as illustrated in Fig. 4(a). Such a spatially localized modulation $\chi(\mathbf{r},t)$ can then provide the necessary $\mathbf{k}$ components to couple different modes. Equation (15), together with the relevant overlap integral, also shows that if $\tilde{\chi}(\omega<0)=0$ (spectrally causal), then the modulation will excite only the modes $n$ for which $\omega_n > \omega_m$. Similarly, if $\tilde{\chi}(\omega>0)=0$ (spectrally anti-causal), then only the modes $n$ with $\omega_n < \omega_m$ will be excited (here we use the term "spectrally anti-causal" in direct analogy to the concept of anti-causality in the temporal domain [52]). Hence, if the time-varying system is designed to be spectrally causal or anti-causal by suitably pairing a temporal modulation of the refractive index with a modulation of the loss coefficient (no gain is required here), we obtain truly unidirectional "propagation" along the synthetic discrete dimension formed by the frequencies of the modes of the resonator. In addition, such a loss-induced unidirectional transport effect is direction-tunable in synthetic space depending on the causality character of the applied modulation. Following this rationale, we designed a temporally modulated ring resonator, and we focused our attention on five of its modes (see middle row of Fig. 4(b) and Supplement 1). The ring resonator was initially excited at the mode $\omega_3$ and three different cases were investigated (see Supplement 1 for details): (A) Spectrally non-causal, (B) spectrally causal, and (C) spectrally anti-causal. We observe in Fig. 4(b) that in case (A) the modulation excites modes bidirectionally along the frequency axis, whereas in cases (B) and (C) the excitation becomes strictly unidirectional, as expected. We note that the physics underlying this form of unidirectional frequency transport is substantially different from other mechanisms that require, for instance, phase modulations with precise, periodic detuning of the modulation frequency from the modal frequency spacing [53]. Here, instead, the unidirectional transport stems from the interplay between refractive-index and loss-coefficient modulation, as can also be noted when comparing cases (B) and (C), which have the same loss profile, but create opposite frequency-transport effects due to the different real part of the modulation. A possible advantage of our method is the fact that the modulation can be applied on demand at any time and does not need to be periodic, avoiding any requirement on the precise control of a periodic resonance-detuning effect (precise knowledge about the modal frequency spacing is actually not needed in our scheme). Thus, the proposed modulation strategy, stemming from the concept of spectral causality, provides a direct and flexible approach to control light propagation along the synthetic frequency dimension, and it might offer new tools to observe and enrich various physical phenomena along synthetic dimensions in systems with structurally lower dimensions.

As a relevant example, one important consequence and immediate application of the unidirectional synthetic transport effects discussed here is the possibility to achieve nonreciprocal frequency conversion without the need for space-time modulations. To this aim, the ring resonator in Fig. 4 may be coupled to an input/output waveguide to excite a certain mode at a certain frequency. Then, the excited mode would be able to couple to higher-frequency modes within the time-modulated structure, whereas the generated higher frequencies would be prohibited from coupling back into the original mode if they are fed back into the system. Hence, nonreciprocal mode/frequency transitions can be achieved without necessarily resorting to space-time, traveling-wave modulations [8] or tandem phase modulators [54]. A potential advantage compared to previously demonstrated Hermitian time-varying nonreciprocal systems is the lack of any coupling oscillations (*i.e.*, the wave does not couple back and forth between modes) thanks to the one-way coupling effect ensured by spectral causality. For certain nonreciprocal structures based on frequency/mode transitions, this fact might relax the strict requirement of terminating the structure at an exact location after which the propagating mode would couple back into its original state [8] and could ease the associated bandwidth limitations [55].

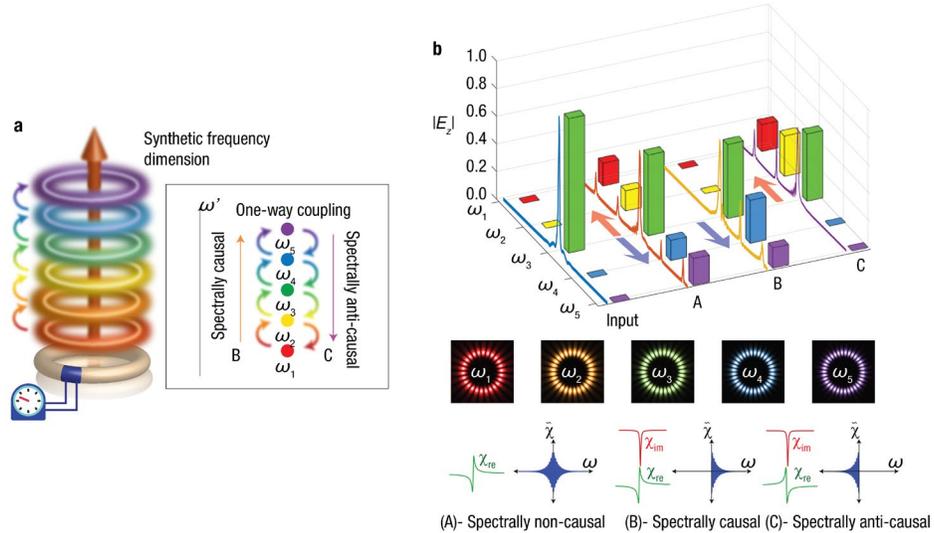

**Fig. 4.** Loss-induced unidirectional transport along a synthetic dimension. (a) Temporal modulation of the material parameters of a resonant cavity (e.g., a ring resonator) can couple different modes at different frequencies. This creates "propagation" along the 'synthetic' dimension formed by the sequence of the frequencies of these resonant modes. The direction of propagation in synthetic space depends on the spectral causality properties of the modulation. (b) Excited frequency components (resonant modes) for an initially excited mode $\omega_3$ in three different scenarios (see lower row): A- spectrally non-causal Hermitian modulation leading to bidirectional frequency transport, B- spectrally causal modulation leading to unidirectional blue-shifting, C- spectrally anti-causal modulation leading to unidirectional red-shifting. Middle row: Spatial electric field amplitude profiles of the resonator modes. Details about the considered resonator geometry, full-wave simulations, and modulation parameters are given in Supplement 1.

## 4. Conclusions

While the scattering response of spatially varying optical systems has been intensively studied over several decades, wave scattering due to time-varying optical properties has received relatively less attention, especially in the context of non-Hermitian (*i.e.* complex) time modulations. Although a few recent works have shown that the notion of PT-symmetric systems can be translated into the temporal realm, no general method existed so far to design broader classes of non-Hermitian temporal modulations with pre-determined scattering properties. In this regard, we believe that the present work provides relevant tools – the concept of spectral causality and the temporal anti-Kramers-Kronig relations – to extend and generalize temporally modulated (meta)materials to the complex domain. More broadly, by emulating the principle of causality, and its implications, in the spectral domain, our findings may open a new landscape of opportunities to control the scattering of waves beyond what is achievable with conventional time-invariant structures and devices. As relevant examples, we have shown that these ideas have direct implications for several applications, such as broadband reflectionless absorption, temporal invisibility and cloaking, and unidirectional/nonreciprocal frequency translation. Incidentally, we note that this notion of spectral causality for time-varying non-Hermitian materials is mathematically related to the concept of single-sideband modulations in communication systems [56], an analogy that suggests the possibility of taking further inspiration from analog communication systems to design even more advanced time-varying metamaterials.

From a practical standpoint, we also note that, although experimental limitations might hinder the realization of very fast modulations in the optical domain, the proposed concepts are general and could potentially be implemented in different frequency regimes with lower modulation speeds. For example, for an experimental demonstration at microwave frequencies, one could implement the required non-Hermitian modulation for the applications discussed here by employing time-varying transmission lines [57] based on dynamically variable capacitances (namely, varactors), variable resistors, active non-Foster elements [58], microwave tunnel diodes [59], etc. Specifically, for the spectrally causal absorber in Section 3.1 one could create a time-varying leaky-wave structure (as was done, for instance, in Ref. [60] with a modulation speed of 900 MHz at around 3-4 GHz) implementing the required spectrally causal modulation to capture an incoming pulse and absorb it with no reflections. To realize a proof of concept of the temporal invisibility effect in Section 3.2, one could similarly employ a time-varying transmission line coupled to input and output coaxial ports (as was done, for instance, in Ref. [61] with a modulation speed of 675 MHz at around 2.5 GHz) and monitor the reflection and transmission coefficients in the presence of a cloaked or non-cloaked temporal perturbation of the transmission-line parameters. For experimental realizations at optical frequencies, one could rely on a composite/layered material platform to modulate the real and imaginary part of the permittivity in different layers with more flexibility, as was recently demonstrated with 2D semiconductor monolayers combined with transition metal dichalcogenides in a ring resonator (where the real and imaginary part of the refractive index can be varied through electrostatic gating) [62],[63]. Such a composite ring resonator structure is a promising platform to demonstrate and study the unidirectional transport effects along a synthetic dimension discussed in Section 3.3, and we are currently exploring the specific requirements for such a demonstration. For example, one could employ a ring resonator with a relatively large radius to reduce the modal frequency spacing and, hence, drastically reduce the requirements on the modulation speed to induce coupling between modes (as done in Ref. [64] but, in our case, with inherently one-way transitions). Moreover, we note that the spectral causality feature is independent of the absolute modulation amplitude. Since Eqs. (7) and (8) can be scaled by an arbitrary factor, only the relative amplitude of the real- and imaginary-part modulations is important to obtain a spectrally causal system. This property could further relax the requirements for the experimental demonstration of these time-varying structures . Finally, due to their generality, the proposed time-modulated systems can also be translated to different domains of wave physics, including in acoustics and elastodynamics, which may provide a particularly fertile ground to test some of the ideas proposed in this paper.

In summary, our findings may open intriguing new opportunities to control, in time and space, the propagation and scattering of waves for applications in various areas of wave physics and engineering.

**Acknowledgments.** F.M. acknowledges support from the Air Force Office of Scientific Research with Grant No. FA9550-19-1-0043 through Dr. Arje Nachman and the National Science Foundation with Grant No. 1741694.Z.H. acknowledges support through the Fulbright Foreign Student Program of the U.S. Department of State.

(†) Online link to Visualization 1: http://tiny.cc/spect_caus_mov1 and Visualization 2: http://tiny.cc/spect_caus_mov2

# Spectral causality and the scattering of waves: supplemental document

ZEKI HAYRAN, AOBO CHEN, AND FRANCESCO MONTICONE *

*School of Electrical and Computer Engineering, Cornell University, Ithaca, New York 14853, USA*
*\* francesco.monticone@cornell.edu*

## 5. Materials and methods

In this work, causal full-wave simulations have been performed using the finite-difference time-domain method through a commercially available software (Lumerical FDTD Solutions, https://www.lumerical.com/products/fdtd/). The time-varying dispersive susceptibility profiles were implemented as custom material plugins by employing the auxiliary differential equation method [1].

In Fig. 2, the input pulse has a center frequency of 530 THz and a bandwidth of 100 THz. The complex susceptibility profile of the medium follows Eq. (9) with the following parameters: $\varepsilon_s = 1$, $\omega_0 = 2\pi500$ THz, $\omega_p = 0.9\,\omega_0$, $\gamma = 0.014\,\omega_0$, $K = 0.0021\,\omega_0^2$, and $t_{\text{offset},1} = -0.14$ ps. In the static case $K$ is set to 0. Both structures are one-dimensional along the $x$- axis and semi-infinite from $x = 0$ to $x \to \infty$.

In Fig. 3, the input pulse has a center frequency of 375 THz and a bandwidth of 22 THz. The complex susceptibility profile of the "cloaked" time-varying medium follows Eq. (10) with the following parameters: $\varepsilon_s = 1$, $f_1 = -1$, $\omega_0 = 2\pi500$ THz, $\omega_p = 0.06\,\omega_0$, $\gamma = 0.002\,\omega_0$, $K = 6.4\times10^{-4}\,\omega_0^2$, $t_{\text{offset},1} = -0.6$ ps, and $t_{\text{offset},2} = -0.7$ ps. The purely Hermitian susceptibility profile of the "uncloaked" time-varying medium follows a lossless Drude dispersion model (with a plasma frequency equal to $\omega_p = 2\pi3.75$ THz) where the high-frequency limit of the susceptibility is modulated in time to obtain the real susceptibility profile in Fig. 3(b) at 375 THz. Additionally, a global high-frequency susceptibility offset of 2 was added (with the simulation parameters re-scaled accordingly) to avoid any numerical instability arising due to negative susceptibilities. In the spectrally non-causal case (in the animation in Supplementary Material) the parameters are the same as in the cloaked case, except $t_{\text{offset},1} = 0.15$ ps. In all three cases the structure is one-dimensional and spatially-invariant along the $x$- axis.

In Fig. 4, the electric field magnitude has been recorded at time $t = 1.5$ ps. The ring resonator has an inner and outer radius equal to 0.8 and 1.0 μm, respectively, and a static relative permittivity $\varepsilon_s = 3$. The perturbed region subtends an angle of 4° with respect to the origin of the resonator. The frequencies of the modes $\omega_1$, $\omega_2$, $\omega_3$, $\omega_4$, and $\omega_5$ are equal to $2\pi404$ THz, $2\pi429$ THz, $2\pi454$ THz, $2\pi479$ THz, and $2\pi504$ THz; respectively. The time-varying complex susceptibility profile of the perturbed region of the ring resonator follows Eq. (9). The parameters for case B are as follows: $\omega_0 = 2\pi500$ THz, $\omega_p = 0.4\,\omega_0$, $\gamma = 0.01\,\omega_0$, $K = 6.4\times10^{-4}\,\omega_0^2$, and $t_{\text{offset}} = -0.4$ ps. The parameters for case C are the same as for case B, except $t_{\text{offset}} = 0.7$ ps. On the other hand, the time-varying purely Hermitian susceptibility profile in case A are the same as the real part of the susceptibility profile in case B, except $\omega_p = 0.25\,\omega_0$. The purely Hermitian susceptibility profile has been modeled as a lossless Drude dispersion model (having a plasma frequency equal to $\omega_p = 2\pi4.54$ THz) with a time-varying high-frequency limit.

## 6. Derivation of the scattering series

The recursive temporal equation given in Eq. (3) (in the main article) can be reformulated as a scattering series in the frequency domain. By Fourier transforming Eq. (3) (in the main article) and noting that the convolution integral in time domain becomes a multiplication in frequency domain, we obtain

$$\tilde{D} = 2\pi D_0 \delta(\omega - \omega_i) + \tilde{G}(\omega)\left[\left(\tilde{\chi} * \left(\omega^2 \tilde{D}\right)\right)(\omega)\right]. \tag{S1}$$

Replacing $\tilde{D}$ on the right-hand side of Eq. (S1) with the full expression of $\tilde{D}$ given by the equation itself yields,

$$\tilde{D} = 2\pi D_0 \delta(\omega - \omega_i) + \tilde{G}(\omega)\left[\left(\tilde{\chi} * \left(\omega^2 \left(2\pi D_0 \delta(\omega - \omega_i) + \tilde{G}(\omega)\left[\left(\tilde{\chi} * \left(\omega^2 \tilde{D}\right)\right)(\omega)\right]\right)\right)\right)(\omega)\right], \tag{S2}$$

$$\tilde{D} = 2\pi D_0 \delta(\omega - \omega_i) + 2\pi \omega_i^2 D_0 \tilde{G}(\omega)\left[(\tilde{\chi} * \delta)(\omega - \omega_i)\right] + \tilde{G}(\omega)\left[\tilde{\chi} * \left[\omega^2 \tilde{G}(\omega)\left[\tilde{\chi} * (\omega^2 \tilde{D})(\omega)\right]\right](\omega)\right]. \tag{S3}$$

The first two terms on the right-hand side of Eq. (S3) correspond to the temporal analogue of the first-order Born approximation in the context of spatial scattering [2]. Such a first-order approximation would be based on a weak-scattering assumption which assumes that the input signal does not undergo a strong perturbation due to its interaction with the medium, and would, therefore, require a very low refractive index contrast (typically on the order of 0.001 [3]). To go beyond such a constraint, we continue the recursive procedure by replacing $\tilde{D}$ on the right-hand side of Eq. (S3) with the full expression of $\tilde{D}$ as in Eq. (S1) to obtain

$$\begin{aligned}\tilde{D} = &\, 2\pi D_0 \delta(\omega - \omega_i) + 2\pi \omega_i^2 D_0 \tilde{G}(\omega)\left[(\tilde{\chi} * \delta)(\omega - \omega_i)\right] + 2\pi \omega_i^2 D_0 \tilde{G}(\omega)\left[\left(\tilde{\chi} * \left[\omega^2 \tilde{G}(\omega)\left[(\tilde{\chi} * \delta)(\omega - \omega_i)\right]\right]\right)(\omega)\right] \\ &+ \tilde{G}(\omega)\left[\left(\tilde{\chi} * \left[\omega^2 \tilde{G}(\omega)\left[\left(\tilde{\chi} * \left[\omega^2 \tilde{G}(\omega)\left[\left(\tilde{\chi} * (\omega^2 \tilde{D})\right)(\omega)\right]\right]\right)(\omega)\right]\right]\right)(\omega)\right],\end{aligned} \tag{S4}$$

where now the first three terms on the right-hand side of Eq. (S3) can be designated as the second-order *temporal* Born approximation. By repeating similar steps, we obtain the temporal scattering series for any order $N$ as in Eqs. (4-6) in the main article.

## 7. Spectral profile of the susceptibility function

Figure S1(a) shows the $\tilde{\chi}(\omega, \omega')$ spectrum of the susceptibility profile given in Eq. (9) (in the main article), where the causal spectrum (*i.e.* $\tilde{\chi}(\omega < 0, \omega') = 0$) can be clearly observed. Moreover, Fig. S1(b) shows the complex susceptibility temporal variation at a specific $\omega'$. The real and imaginary part of the *t*-time variation of the susceptibility approximately satisfies the anti-Kramers-Kronig relations, while their $\omega$-spectrum has a nearly causal behavior as seen in Fig. S1(c).

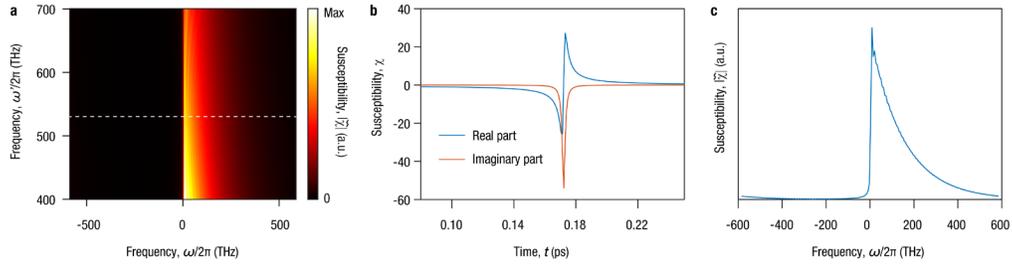

**Fig. 1.** Susceptibility spectral profile. (a) The $\tilde{\chi}(\omega, \omega')$ spectrum of the susceptibility profile given in Eq. (9) (in the main article). The temporal variation of the susceptibility profile along the $t$-axis (b) and the magnitude of its Fourier spectrum along the $\omega$-axis (c) at $\omega'/2\pi$ = 530 THz (denoted with the white dashed line in (a)). The parameters are the same as in Fig. 2 (in the main article).